\documentclass[letterpaper, 10 pt, conference]{ieeeconf}  

\IEEEoverridecommandlockouts                              
\usepackage{graphics} 
\usepackage{epsfig} 
\usepackage{mathptmx} 
\usepackage{times} 
\usepackage{amsmath} 
\usepackage{amssymb}  
\usepackage{hyperref}
\usepackage{subcaption}
\usepackage{multirow}
\usepackage{bm}
\title{Comparing Parameterizations and Objective Functions for Maximizing the Volume of Zonotopic Invariant Sets}

\author{Chenliang Zhou, Heejin Ahn and Ian M. Mitchell
\thanks{The authors are with the Department of Computer Science,
        University of British Columbia {\tt\small chenliang.zhou@qq.com, hjahn@cs.ubc.ca, ian.mitchell@ubc.ca}. Research supported by the National Science and Engineering Research Council of Canada (NSERC) Discovery Grant \#4543.}%
}

\usepackage[T1]{fontenc}

\usepackage{todonotes}

\hypersetup{pdfpagemode=UseNone}

\usepackage{paralist}

\graphicspath{{Figures/}}

\newcommand{\matlab}{\textsc{Matlab}}

\newcommand{\R}{\mathbb{R}}
\newcommand{\bigO}{\mathcal{O}}

\newcommand{\defined}{\triangleq}

\newcommand{\reachOp}{\mathsf{R}}
\newcommand{\reachSet}[2]{\reachOp\left({#1},{#2}\right)}

\newcommand{\centvecOp}{c}
\newcommand{\centvec}[1]{\centvecelem{}{#1}}
\newcommand{\centvecelem}[2]{{\centvecOp}_{#1}\left({#2}\right)}

\newcommand{\genmatOp}{\mat{G}}
\newcommand{\genmat}[1]{\genmatOp\left({#1}\right)}
\newcommand{\countvecOp}{p}
\newcommand{\countvec}[1]{\countvecOp\left({#1}\right)}
\newcommand{\dimvecOp}{d}
\newcommand{\dimvec}[1]{\dimvecOp\left({#1}\right)}
\newcommand{\scalevec}{\gamma}
\newcommand{\scalemat}{\Gamma}

\newcommand{\ones}[1]{\mathbf{1}_{#1}}
\newcommand{\zeros}[1]{\mathbf{0}_{#1}}
\newcommand{\mat}[1]{\mathrm{#1}}
\newcommand{\set}[1]{\mathsf{#1}}
\newcommand{\com}[1]{\texttt{#1}}

\newcommand{\bma}{\begin{bmatrix}}
\newcommand{\ema}{\end{bmatrix}}

\long\def\ignore#1{}
\DeclareMathOperator{\rank}{rank}
\DeclareMathOperator{\vol}{vol}
\DeclareMathOperator{\trace}{tr}
\DeclareMathOperator{\diag}{diag}
\DeclareMathOperator{\etr}{etr}

\newtheorem{theorem}{Theorem}
\newtheorem{proposition}[theorem]{Proposition}
\newtheorem{definition}[theorem]{Definition}

\newcommand{\todoIM}[1]{\todo[inline,color=green,author=IM]{#1}}

\begin{document}
\maketitle
\thispagestyle{empty}
\pagestyle{empty}

\begin{abstract}
In formal safety verification, many proposed algorithms use parametric set representations and convert the computation of the relevant sets into an optimization problem; consequently, the choice of parameterization and objective function have a significant impact on the efficiency and accuracy of the resulting computation.  In particular, recent papers have explored the use of zonotope set representations for various types of invariant sets.  In this paper we collect two zonotope parameterizations that are numerically well-behaved and demonstrate that the volume of the corresponding zonotopes is log-concave in the parameters.  We then experimentally explore the use of these two parameterizations in an algorithm for computing the maximum volume zonotope invariant under affine dynamics within a specified box constraint over a finite horizon.  The true volume of the zonotopes is used as an objective function, along with two alternative heuristics that are faster to compute.  We conclude that the heuristics are much faster in practice, although the relative quality of their results declines as the dimension of the problem increases; however, our conclusions are only preliminary due to so-far limited availability of compute resources.
\end{abstract}

\section{Introduction}

A common approach to safety analysis and/or synthesis involves the computation of reachable, invariant, viable or discriminating sets.  Nonparametric representations of such sets, such as implicit surface functions sampled on a grid~\cite{mitchell2005time}, are general but scale poorly with dimension; consequently, much work has focused on parametric representations including polytopes \cite{frehse2011spaceex, trodden2016one, zhang2011unbounded}, support functions \cite{frehse2011spaceex, leguernic2009reachability}, and ellipsoids \cite{mitchell2016ensuring, li2006estimating, mitchell2015improved} because of  their computational efficiency on some of the common set operations. Zonotopes, a class of polytopes, have attracted particular attention recently because of the flexibility and efficiency of their encoding, as well as the fact that they are closed under affine transformation and Minkowski sum~\cite{leguernic2009reachability,althoff2016combining}.  Starting with~\cite{girard2005reachability} for reachable sets, they have been used in a wide variety of set computations, particularly for systems with linear or affine dynamics.

While many algorithms construct their sets by stepping in time, the advent of powerful linear, convex and general nonlinear optimization tools has lead to a category of algorithms which seek to construct the sets through optimization. For example, in \cite{korda2014convex}, they presented approaches to approximate viability kernels based on the computation of the region of attraction and then converted the problem to an infinite-dimensional linear programming problem in the cone of nonnegative occupation measures. Another use of convex optimization is presented in \cite{liu2019full} to approximate polytopic discriminating kernels and the corresponding linear state feedback control law in a linear discrete-time system. The authors of \cite{yu2018control} consider the computation of discriminating sets and the corresponding control law for systems with additive and bounded disturbances by constructing linear matrix inequalities optimization problem.  Choice of efficient parameterization is the focus of \cite{gupta2019computation}, which considers the computation of discriminating sets for linear discrete-time feedback systems where they assume the discriminating sets are of low-complexity, namely symmetric around the origin and described by the same number of affine inequalities as twice the dimension of the state variables.

In the context of optimization, the question of parameterization is critical to efficient solution, since optimization cost is typically superlinear in the number of decision variables and constraints.  In particular, two zonotope parameterizations have been used recently in the literature exploring invariant sets.  Template zonotopes are used in~\cite{admioolam2017templatezonotope}: Zonotopes whose generators are fixed but the scaling of those generators can be adjusted.  Parallelotopes (a subclass of zonotopes) are used in~\cite{sadraddini2019sampling}: Zonotopes whose generator matrix is upper triangular with positive entries along the diagonal.  Both parameterizations can easily ensure that the zonotope volume does not collapse to zero, which is much more difficult to achieve with a general dense generator matrix.

Our contributions in this paper are to:
\begin{compactitem}
    \item Collect the two zonotope parameterizations from~\cite{admioolam2017templatezonotope,sadraddini2019sampling} and show for both that the volume of the zonotope is a log-concave function of the free parameters and is hence amenable to efficient maximization.
    \item Compare experimentally the use of these two parameterizations in an algorithm for computing a maximum volume zonotope invariant under affine, input-free dynamics in a given box constraint over finite horizon from~\cite{mitchell2019invariant}.  For one of the parameterizations the true volume is computationally expensive to evaluate, so we also compare two heuristic objective functions for that parameterization.
\end{compactitem}

\section{PRELIMINARIES}

We define the invariance problem that we seek to solve, and the zonotope set representation and parameterizations that we use to solve it.

\subsection{Invariant Sets}

In this paper we consider discrete-time, time-invariant affine dynamics
\begin{equation} \label{eqn:dynamic_system}
x(t+1) = \mat Ax(t) + w,
\end{equation}
where $x(t) \in \R^{d_x}$ represents the state at time $t \in \mathbb{N} = \{0, 1, 2, \ldots\}$, $\mat A \in \R^{d_x \times d_x}$ is the evolution matrix, and $w \in \R^{d_x}$ is the constant drift.

Given a system \eqref{eqn:dynamic_system}, a box constraint set 
\[
  \set X = \{x \in \R^{d_x} \mid \underline{x} \leq x \leq \overline{x} \},
\]
and a time $T \in \mathbb{N}$, we would like to find a \emph{finite horizon invariant set}.
\begin{definition}
A set $\set I$ is invariant in $\set X$ for horizon $[0,T]$ if for all $x(0) \in \set I$ and for all $t \in [0, T]$, $x(t) \in \set X$.
\end{definition}

There are contexts in which one might seek invariant sets with maximal or minimal volume, but in the remainder of this paper we focus on the former.

Define the \emph{forward reach set} of some specified set $\set S$ as
\begin{equation} \label{e:reach-defn}
  \reachSet{t}{\set S} \defined
      \left\{ x(t) \in \R^{d_x} \,\left|\, x(0) \in \set S 
      \right. \right\}.
\end{equation}
Note that the reach set is defined at a single time rather than over a time interval, and the set $\set S$ is an initial condition rather than a constraint.  We will use the reach set of $\set I$ to constrain $\set I$ to be invariant.
\begin{proposition} \label{t:invariance-from-reach}
A set $\set I$ is invariant in $\set X$ over horizon $[0,T]$ if for all $t \in [0, T]$, $\reachSet{t}{\set I} \subseteq \set X$.
\end{proposition}

\subsection{Zonotope Representation}

We choose to use the compact zonotope to represent our invariant sets. A \emph{zonotope} $\set Z \subset \R^{\dimvec{\set Z}}$ can be characterized by its center $\centvec{\set Z} \in \R^{\dimvec{\set Z}}$ and generator matrix $\genmat{\set Z} \in \R^{\dimvec{\set Z} \times \countvec{\set Z}}$ as the set
\[
  \set Z = \left\{ \centvec{\set Z} + \genmat{\set Z} \lambda
    \, \left| \, -\ones{\countvec{\set Z}} \leq \lambda \leq +\ones{\countvec{\set Z}} 
    \right. \right\}.
\]
For a zonotope $\set Z$ with center $\centvecOp$ and generator matrix $\genmatOp$, we will sometimes write $\langle \centvecOp \mid \genmatOp \rangle$.  If the generator matrix $\genmatOp \in \R^{\dimvecOp \times \countvecOp}$ is such that $\countvecOp \geq \dimvecOp$ and $\rank(\genmatOp) = \dimvecOp$---in other words, that $\genmatOp$ is short and fat and has full column rank---then the corresponding zonotope $\langle \centvecOp \mid \genmatOp \rangle \subset \R^{\dimvecOp}$ is full dimensional.

For a zonotope with generator matrix $\genmatOp \in \R^{\dimvecOp \times \countvecOp}$ and $\rank(\genmatOp) = r$, its $r$-dimensional volume can be computed by the formula~\cite{gover2010determinants}
\begin{equation} \label{eqn:zonotope-volume}
    \vol_r(\genmatOp) = 
      2^r \sum_{\{j_i\}_{i=1}^r} 
      \sqrt{\det \left[(\genmatOp^{\{j_i\}})^T (\genmatOp^{\{j_i\}}) \right]},
\end{equation}
where the summation is over the set of $r$ indexes $\{ j_i \}_{i=1}^r$ such that $1 \leq j_1 < j_2 < \cdots < j_r \leq \countvecOp$ and $\genmatOp^{\{j_i\}} \in \R^{\dimvecOp \times r}$ is the submatrix of $\genmatOp$ consisting only of columns $j_1, j_2, \ldots, j_r$.  Note that if we vary the entries of a generator matrix $\genmatOp$, the volume formula~\eqref{eqn:zonotope-volume} discretely changes form (fewer columns are chosen) when the rank drops.  In order to avoid such a complication, we choose constraints on the parameterizations of our zonotopes to ensure their generator matrices are of full rank.

In our search for invariant sets, we will consider two parameterizations of zonotopes.  The distinction between the two lies in the choice of generator matrix; in both cases we allow the center vector to be chosen arbitrarily.  

\emph{Upper Triangular Positive Definite} (UTPD): Following~\cite{sadraddini2019sampling} we constrain the generator matrix to be upper triangular with strictly positive entries along the diagonal; consequently, the generator matrix must be square and full rank, and the zonotope is a full dimensional parallelotope.  For a zonotope $\set Z \subset \R^{d_x}$, the corresponding UTPD matrix $\genmat{\set Z}$ will have $\tfrac{1}{2} d_x (d_x+1)$ free variables and $d_x$ positivity constraints.

\emph{Scaled Fixed Generators} (SFG): Following~\cite{admioolam2017templatezonotope} but restricting entries to be real valued as in~\cite{mitchell2019invariant} we constrain the generator matrix to take the form $\genmatOp \scalemat$ where $\genmatOp \in \R^{\dimvecOp \times \countvecOp}$ is fixed and $\scalemat = \diag(\scalevec) \in \R^{\countvecOp \times \countvecOp}$ is the diagonal scaling matrix with vector $\scalevec \in \R^{\countvecOp}$ along its diagonal.  If we choose $\genmatOp$ to have full rank, then the constraint $\scalevec > 0$ ensures that the product $\genmatOp \scalemat$ is also full rank.  The corresponding zonotope is a full dimensional (real) template zonotope in the sense that the directions of the generators are fixed but their scaling is not.  This parameterization has $\countvecOp$ free parameters (the entries of $\scalevec$) and $\countvecOp$ positivity constraints.

The \emph{order} of a zonotope $\set Z$ is defined as $\frac{\countvec{\set Z}}{\dimvec{\set Z}}$; consequently, the UTPD parameterization is always order 1 while the SFG parameterization has the same order as the predefined $\genmat{\set Z}$.

\section{Invariant Set Construction as a Convex Optimization}

For dynamics~\eqref{eqn:dynamic_system} it is straightforward to show that the reach set of an initial zonotope $\set I$ is itself a zonotope characterized by
  \begin{equation} \label{e:zonotope-no-input-evolution}
    \begin{aligned}
      \centvec{\reachSet{t}{\set I}} &= \mat A^t \centvec{\set I}
        + \sum_{s=0}^{t-1} \mat A^{t-1-s} w \\
      \countvec{\reachSet{t}{\set I}} &= \countvec{\set I} \\
      \genmat{\reachSet{t}{\set I}} &= A^t \genmat{\set I} \Gamma,
    \end{aligned}
  \end{equation}
Using Proposition~\ref{t:invariance-from-reach} and~\eqref{e:zonotope-no-input-evolution} we arrive at a set of constraints on $\set I$ to achieve invariance.
\begin{proposition}\label{t:invariant-constraints}
\cite[Proposition~4.2]{mitchell2019invariant}
Zonotope $\set I$ is invariant in $\set X$ over horizon $[0,T]$ if
  \begin{equation} \label{e:invariant-no-input-box-containment}
    \begin{aligned}
       \mat A^t \centvec{\set I} + \sum_{s=0}^{t-1} \mat A^{t-1-s} w
         - \left|\mat A^t \genmat{\set I}\right|
        &\geq \underline x, \\
       \mat A^t \centvec{\set I} + \sum_{s=0}^{t-1} \mat A^{t-1-s} w
         + \left|\mat A^t \genmat{\set I}\right|
        &\leq \overline x. \\
    \end{aligned}
  \end{equation}
  for all $t = 0, \ldots, T$.
\end{proposition}

These constraints are linear in the entries of $\centvec{\set I}$ and convex in the entries of $\genmat{\set I}$.  In fact, it can be shown that the constraints are linear in any components of $\genmat{\set I}$ with known sign; for example, the diagonal entries of $\genmat{\set I}$ in the UTPD parameterization or the entries of the scaling vector $\scalevec$ in the SFG parameterization.

Using Proposition~\ref{t:invariant-constraints}, we can construct an optimization to find the largest invariant zonotope within our parameterized set of zonotopes.
\begin{equation} \label{e:invariant-optimization-no-input}
  \begin{aligned}
  \max_{\centvec{\set I}, \genmat{\set I}} \; \vol(\genmat{\set I}) \\
  \text{such that } & \text{the rank constraint on } \genmat{\set I} \text{ is met} \\
  \text{and } & \text{\eqref{e:invariant-no-input-box-containment} holds }
                \forall t \in 0, \ldots, T\\
  \end{aligned}
\end{equation}

For the UTPD and SFG parameterizations of $\set I$, the rank constraint translates into linear inequalities, so~\eqref{e:invariant-optimization-no-input} will be a convex optimization if the objective function is convex.  We explore the implications for each of the parameterizations in the next subsections.

\subsection{Objective Functions for the UTPD Parameterization}

The UTPD parameterization uses a square, upper-triangular generator matrix, so the sum in~\eqref{eqn:zonotope-volume} collapses to a single term
\begin{equation} \label{e:zonotope-volume-UTPD}
  \begin{aligned}
    \vol_{d_x}(\genmatOp) &= 2^{d_x} \sqrt{\det\left[\genmatOp^T \genmatOp\right]},\\
      &= 2^{d_x} \sqrt{\prod_{i=0}^{d_x} \genmatOp_{i,i}^2},\\
      &= 2^{d_x} \prod_{i=0}^{d_x} \genmatOp_{i,i},
  \end{aligned}
\end{equation}
where $\genmatOp_{i,i}$ is the $i^{th}$ diagonal element of $\genmatOp$ and the final equality holds because the rank constraint for UTPD specifies $\genmatOp_{i,i} > 0$.

\begin{proposition}
For the UTPD parameterization, the zonotope volume is log-concave (actually log-linear) in the free variables (the entries of the upper triangular generator matrix).
\end{proposition}

\begin{proof}
The (identity) function $\genmatOp_{i,i}$ is log-concave on $\genmatOp_{i,i} > 0$ for each $i$, and the product of log-concave functions is log-concave.
\end{proof}

For UTPD the cost of evaluating the convex constraints~\eqref{e:invariant-no-input-box-containment} is $\bigO(T d_x^3)$, dominated by the cost of evaluating the $|\mat A^t \genmat{\set I}|$ term.  The cost of evaluating the objective function~\eqref{e:zonotope-volume-UTPD} is $\bigO(d_x)$.  

Although~\cite{sadraddini2019sampling} proposes a linear objective function for maximizing the volume of their zonotopes (allowing them to keep the continuous portion of their optimization linear), we do not further explore their objective function here because the true volume formula~\eqref{e:zonotope-volume-UTPD} is log-concave and so cheap to evaluate.

\ignore{
We also consider the \emph{extended trace} objective function used in~\cite{sadraddini2019sampling}
\begin{equation} \label{e:etr-defn}
    \etr(\genmatOp) =\sum_{t=0}^T \trace(\mat A^t \genmatOp) +  \beta \epsilon T d_x,
\end{equation}
where
\begin{itemize}
    \item $\mat A$ is the evolution matrix in \eqref{eqn:dynamic_system} and $T$ is the horizon.
    \item $\trace(\genmatOp) = \sum_{i=0}^{d_x} w_i \genmatOp_{i,i}$ is the weighted trace of $\genmatOp$.  For our experiments, we chose uniform weighing $w_i=1$ for all $i$.
    \item The full rank of $\genmatOp$ is ensured by adding $\genmatOp{i,i} \geq \frac{\epsilon}{w_i}$ into the set of constraints as a regularization.  \todoIM{We choose $\epsilon=1$?}.
    \item $\beta$ is a constant for weighing the regularization term. For our experiments, we chose $\beta = 100$.
\end{itemize}
}

\subsection{Objective Functions for the SFG Parameterization}

The SFG parameterization multiplies a fixed rectangular generator matrix $\genmatOp \in \R^{d_x \times \countvecOp}$ by a positive scaling factor $\scalemat = \diag(\scalevec)$, so the summation in~\eqref{eqn:zonotope-volume} is still present but can be simplified.
\begin{equation} \label{e:zonotope-volume-SFG}
  \begin{aligned}
    \vol_{d_x}(\genmatOp) 
      &= 2^{d_x} \sum_{\{j_i\}_{i=1}^{d_x}} 
         \sqrt{\det \left[(\genmatOp^{\{j_i\}} \scalemat^{\{j_i\}})^T 
                          (\genmatOp^{\{j_i\}} \scalemat^{\{j_i\}}) \right]}, \\
      &= 2^{d_x} \sum_{\{j_i\}_{i=1}^{d_x}} 
         \sqrt{\det\left[(\genmatOp^{\{j_i\}})^T (\genmatOp^{\{j_i\}}) \right]
               \prod_{i=1}^{d_x} \scalevec_{j_i}^2}, \\
      &= \sum_{\{j_i\}_{i=1}^{d_x}} 
         \left( \vol(\genmatOp^{\{j_i\}}) \prod_{i=1}^{d_x} \scalevec_{j_i} \right),
  \end{aligned}
\end{equation}
where
\[
  \vol(\genmatOp^{\{j_i\}}) 
    = 2^{d_x} \sqrt{\det\left[(\genmatOp^{\{j_i\}})^T (\genmatOp^{\{j_i\}}) \right]}
\]
is the volume of the parallelotope with (square) generator matrix $\genmatOp^{\{j_i\}}$, and is hence a fixed weight in~\eqref{e:zonotope-volume-SFG} once the generator matrix $\genmatOp$ is chosen.  Note that unlike the (multiple) summation over the multiple indexes $\{j_i\}_{i=1}^{d_x}$, the (single) product in~\eqref{e:zonotope-volume-SFG} is over the single index $i$ and multiplies together the $j_i^{th}$ elements of the scaling vector $\scalevec$.

\begin{proposition}
For the SFG parameterization, the zonotope volume is log-concave in the free variables (the entries of the scaling vector $\scalevec$).
\end{proposition}

\begin{proof}
Assume without loss of generality that zonotope $\set Z$ is centered at the origin, so $\centvec{\set Z} = \zeros{d_x}$.  In the SFG parameterization,
\[
  \begin{aligned}
  \set Z 
    &= \left\{ \genmat{\set Z} \scalemat \lambda
        \, \left| \, -\ones{\countvec{\set Z}} \leq \lambda \leq +\ones{\countvec{\set Z}} 
        \right. \right\}, \\
    &= \left\{ \genmat{\set Z} \lambda
        \, \left| \, -\scalevec \leq \lambda \leq +\scalevec 
        \right. \right\}. \\
  \end{aligned}
\]
Therefore, the indicator function for $\set Z$ can be written as
\[
  \Psi(x,\scalevec) = 
    \begin{cases}
    1, &\exists \lambda \text{ such that } \left\{ \begin{gathered} \genmat{\set Z} \lambda = x \\ -\scalevec \leq \lambda \leq +\scalevec \end{gathered} \right\}; \\
    0, &\text{otherwise}.
    \end{cases}
\]
Because we require $\genmat{\set Z}) \scalemat$ to have full rank, there always exists (many) $\lambda \in \R^{\countvec{\set Z}}$ such that $\genmat{\set Z}\lambda = x$, but if $x \notin \set Z$ then the constraints $-\scalevec \leq \lambda \leq +\scalevec$ cannot be satisfied.

Following~\cite[Example~3.44]{boyd2004convex} it can be shown that $\Psi(x,\scalevec)$ is log-concave in its arguments; thus,
\[
    \vol(\set Z) = \int \Psi(x, \scalevec) \, dx
\]
is log-concave as a function of $\scalevec$.
\end{proof}

For SFG the cost of evaluating the convex constraints~\eqref{e:invariant-no-input-box-containment} is the same as for UTPD: $\bigO(T d_x^3)$.  However, there are 
\[
  \begin{aligned}
    {\countvecOp \choose d_x} &= \frac{\countvecOp!}{d_x!(\countvecOp-d_x)!} \\
      &= \bigO\left(\countvecOp^{d_x}\right) \text{ for fixed } d_x
  \end{aligned}
\]
terms in~\eqref{e:zonotope-volume-SFG} each of which is $\bigO(d_x)$ to evaluate (assuming that the constant weights $\vol(\genmatOp^{\{j_i\}})$ are precomputed); consequently, the cost of evaluating~\eqref{e:zonotope-volume-SFG} grows very fast for higher order zonotopes.

In order to reduce the computational cost of the objective function for the SFG parameterization, we propose two alternative heuristics:
\begin{compactitem}
  \item \emph{Sum of scalars}: Following the positive results reported in~\cite{mitchell2019invariant}, use
  \begin{equation} \label{e:ss-defn}
    \|\countvecOp\|_1 = \sum_{i=1}^\countvecOp \scalevec_i.
  \end{equation}
  This choice is linear in $\scalevec$ and $\bigO(\countvecOp)$ to evaluate.
  \item \emph{Log sum of scalars}: Following the observation that the true volume~\eqref{e:zonotope-volume-SFG} is the sum of weighted products of subsets of the entries of $\scalevec$, use
  \begin{equation} \label{e:slgs-defn}
    \prod_{i=1}^\countvecOp \scalevec_i.
  \end{equation}
  This choice is log-linear in $\scalevec$ and $\bigO(\countvecOp)$ to evaluate.
\end{compactitem}

\section{Experimental Comparison of Parameterizations and Objective Functions}
\label{sec:Comparison of Objective Functions for Invariant Set Algorithm}

In this section we experimentally compare the two parameterizations and three objective functions discussed in the previous section.  The experiments were run on a MacBook Pro laptop with a 2.5~GHz quad-core Intel Core~i7 processor and 16~GB of 1600~MHz DDR3 RAM under macOS Catalina (version~10.15.3). We used the Julia programming language~\cite{bezanson2017julia} with the JuMP package~\cite{dunning2017jump} to formulate our optimizations and the Ipopt solver~\cite{wachter2006implementation} on the backend.

\subsection{Procedure}
In every case we seek to solve the optimization~\eqref{e:invariant-optimization-no-input} and thereby compute a maximum volume zonotope $\set I$ which is invariant in $\set X$ over horizon $T$.  The four cases we examine are:
\begin{compactitem}
\item SFG+ss: SFG parameterization with the sum of scalars heuristic objective function~\eqref{e:ss-defn}.
\item SFG+slgs: SFG parameterization with the product of scalars heuristic objective function~\eqref{e:slgs-defn}.  Because this heuristic is log-linear, in practice we use the objective $\log \prod_i \scalevec_i = \sum_i \log \scalevec_i$ (the sum of logs of scalars).
\item SFG+lgv: SFG parameterization with the true zonotope volume objective function~\eqref{e:zonotope-volume-SFG}.  Because this function is log-concave, we maximize the log of the volume.
\item UTPD+lgv: UTPD parameterization with the true zonotope volume objective function~\eqref{e:zonotope-volume-UTPD}.  Because this function is log-concave, we maximize the log of the volume.
\end{compactitem}

\begin{table}
\caption{Experimental combinations of $d_x$ and $p$}
\label{table:combination_d_p_objective_compare}
\begin{center}
\begin{tabular}{rrrr}
\hline
$d_x$ & $\countvecOp$ & trials & ${\countvecOp \choose d_x}$ \\ \hline
3 & 3 & 2500 & 1\\
3 & 6 & 2500 & 20\\
3 & 8 & 2500 & 56\\
6 & 10 & 2500 & 210 \\
8 & 13 & 100 & 1287 \\
10 & 14 & 100 & 1001\\
12 & 15 & 100 & 455 \\
15 & 16 & 100 & 16 \\ \hline
\end{tabular}
\end{center}
\end{table}

We ran our comparison experiment on eight different combinations of $(d_x,\countvecOp)$ as listed in Table~\ref{table:combination_d_p_objective_compare} in order to explore the impact of the number of dimensions $d_x$ and the number of generators $\countvecOp$ on the performance of different objective functions.  We remind the reader that $\countvecOp > d_x$ is only meaningful in the SFG parameterization, where the fixed generator matrix $\genmatOp \in \R^{d_x \times \countvecOp}$; in the UTPD parameterization the number of generators is always $d_x$.

In each trial, we generated a random stable dynamics matrix $\mat A$ in~\eqref{eqn:dynamic_system} using a Julia reimplementation of \matlab's \com{rss} command (with discrete time step size $0.2$).  For the SFG parameterization we filled the first $d_x$ columns of the fixed generator matrix with the identity matrix of size $d_x$, and any remaining columns for $\countvecOp > d_x$ were filled with random unit vectors (different vectors for each trial).  The constraint set was chosen as $\set X = \{x \mid -\ones{d_x} \leq x \leq +\ones{d_x} \}$ and the horizon $T = 30$.

\subsection{Results}
\begin{table}
\caption{Average Optimal Volumes}
\label{table:average_optimal_compare}
\begin{center}
\begin{tabular}{c|ccc|c}
\hline
 & \multicolumn{3}{|c|}{SFG} & UTPD\\
($d_x$,$\countvecOp$) & ss & slgs & lgv & lgv \\ \hline
(3, 3) & 6.29 & 6.33 & 6.33 & 6.67\\
(3, 6) & 6.60 & 5.47 & 6.78 & " \\
(3, 8) & 6.86 & 5.32 & 7.07 & " \\
(6, 10) & 17.51 & 15.28 & 21.71 & 22.80\\
(8, 13) & 15.71 & 18.81 & 26.22 & 29.30 \\
(10, 14) & 25.88 & 29.28 & 41.36 & 55.73\\
(12, 15) & 22.33 & 31.40 & 39.03 & 84.39\\
(15, 16) & 17.82 & 23.17 & 24.59 & 106.18\\ \hline
\end{tabular}
\end{center}
\end{table}

\begin{table}
\caption{Average Runtimes (seconds)}
\label{table:average_runtime_compare}
\begin{center}
\begin{tabular}{c|ccc|c}
\hline
 & \multicolumn{3}{|c|}{SFG} & UTPD\\
 ($d_x$,$\countvecOp$) & ss & slgs & lgv & lgv\\ \hline
(3, 3) & 0.01 & 0.02 & 0.04 & 0.22\\
(3, 6) & 0.02 & 0.02 & 0.06 &  " \\
(3, 8) & 0.02 & 0.02 & 0.08 &  " \\
(6, 10) & 0.04 & 0.04 & 0.36 & 1.05\\
(8, 13) & 0.07 & 0.07 & 3.65 & 2.80\\
(10, 14) & 0.09 & 0.09 & 5.93 & 5.83\\
(12, 15) & 0.13 & 0.13 & 5.82 & 10.75\\
(15, 16) & 0.22 & 0.20 & 1.19 & 37.42\\ \hline
\end{tabular}
\end{center}
\end{table}

\begin{figure*}[tb]
\centering
\begin{subfigure}{.2\textwidth}
  \centering
  \includegraphics[width=\linewidth]{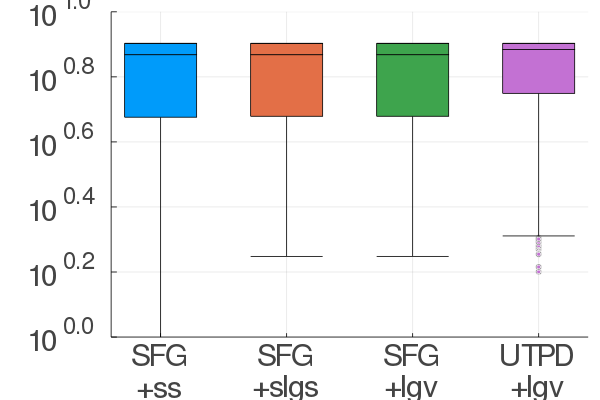}
  \caption{$d_x=3,p=3$}
\end{subfigure}
\hfill
\begin{subfigure}{.2\textwidth}
  \centering
  \includegraphics[width=\linewidth]{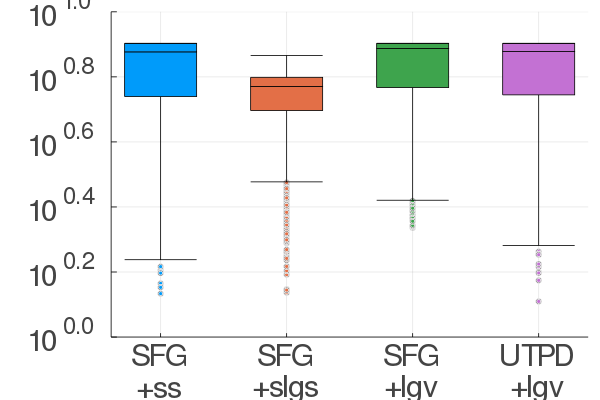}
  \caption{$d_x=3,p=6$}
\end{subfigure}
\hfill
\begin{subfigure}{.2\textwidth}
  \centering
  \includegraphics[width=\linewidth]{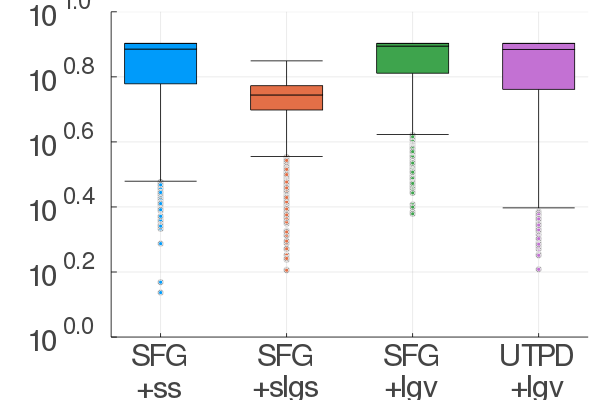}
  \caption{$d_x=3,p=8$}
\end{subfigure}
\hfill
\begin{subfigure}{.2\textwidth}
  \centering
  \includegraphics[width=\linewidth]{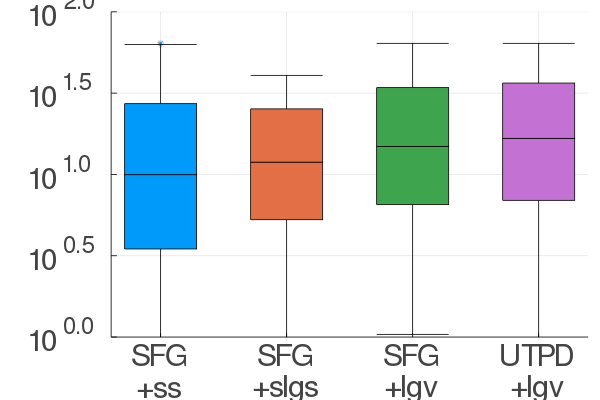}
  \caption{$d_x=6,p=10$}
\end{subfigure}

\begin{subfigure}{.19\textwidth}
  \centering
  \includegraphics[width=\linewidth]{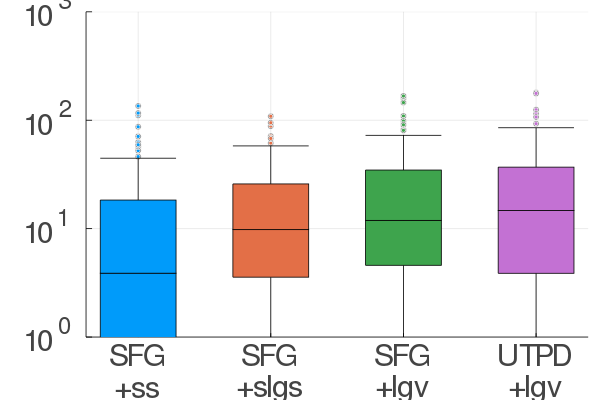}
  \caption{$d_x=8,p=13$}
\end{subfigure}
\hfill
\begin{subfigure}{.19\textwidth}
  \centering
  \includegraphics[width=\linewidth]{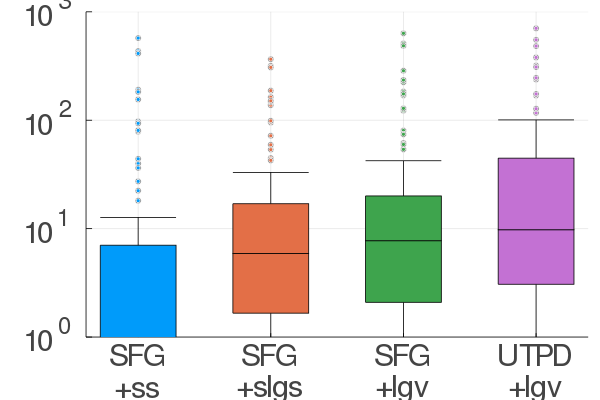}
  \caption{$d_x=10,p=14$}
\end{subfigure}
\hfill
\begin{subfigure}{.19\textwidth}
  \centering
  \includegraphics[width=\linewidth]{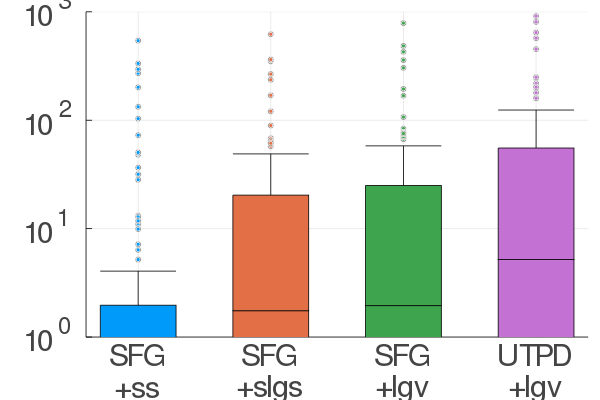}
  \caption{$d_x=12,p=15$}
\end{subfigure}
\hfill
\begin{subfigure}{.19\textwidth}
  \centering
  \includegraphics[width=\linewidth]{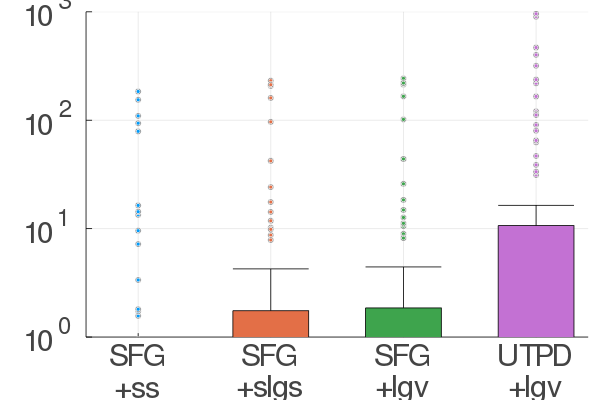}
  \caption{$d_x=15,p=16$}
\end{subfigure}
\caption{Distribution of Optimal Zonotope Volumes}
\label{fig:boxplots_optval}
\end{figure*}

\begin{figure*}[tb]
\centering
\begin{subfigure}{.2\textwidth}
  \centering
  \includegraphics[width=\linewidth]{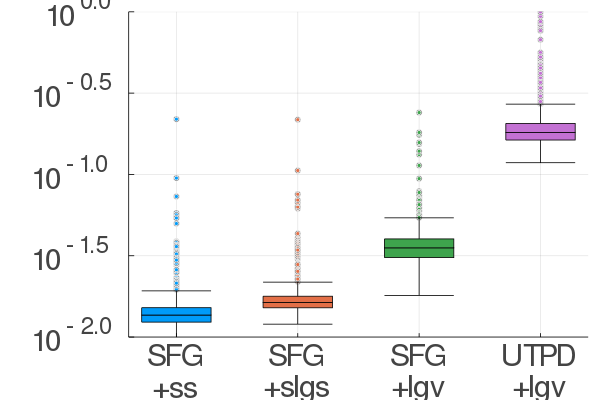}
  \caption{$d_x=3,p=3$}
\end{subfigure}
\hfill
\begin{subfigure}{.2\textwidth}
  \centering
  \includegraphics[width=\linewidth]{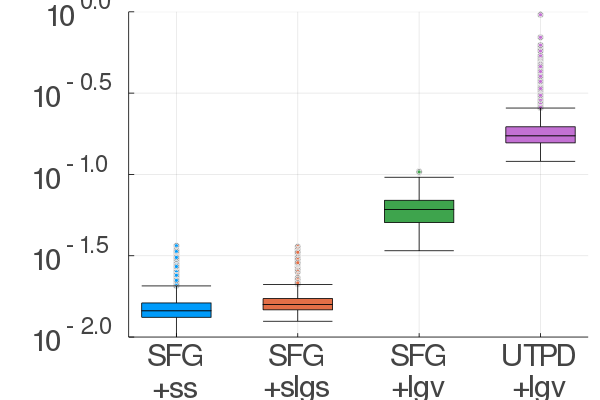}
  \caption{$d_x=3,p=6$}
\end{subfigure}
\hfill
\begin{subfigure}{.2\textwidth}
  \centering
  \includegraphics[width=\linewidth]{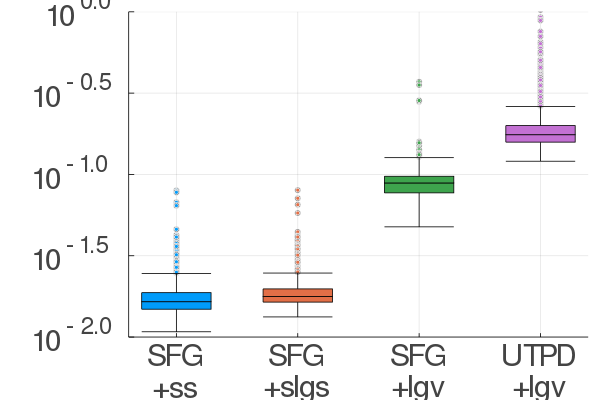}
  \caption{$d_x=3,p=8$}
\end{subfigure}
\hfill
\begin{subfigure}{.2\textwidth}
  \centering
  \includegraphics[width=\linewidth]{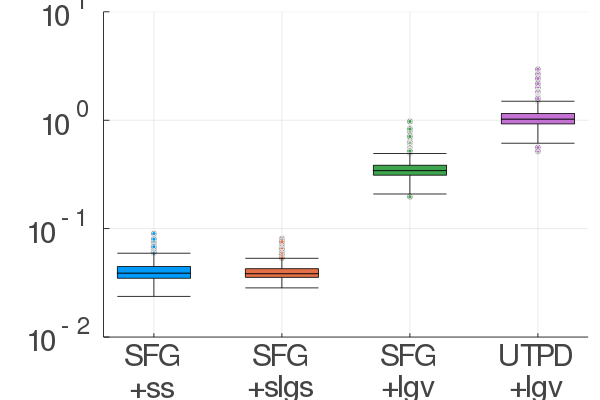}
  \caption{$d_x=6,p=10$}
\end{subfigure}

\begin{subfigure}{.19\textwidth}
  \centering
  \includegraphics[width=\linewidth]{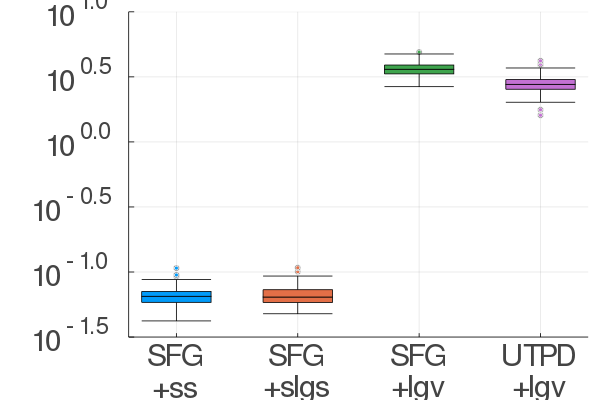}
  \caption{$d_x=8,p=13$}
\end{subfigure}
\hfill
\begin{subfigure}{.19\textwidth}
  \centering
  \includegraphics[width=\linewidth]{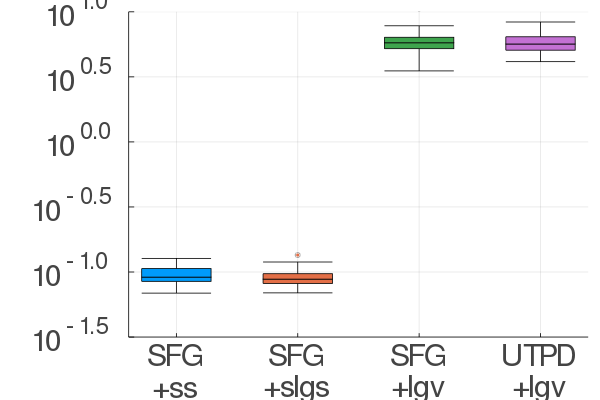}
  \caption{$d_x=10,p=14$}
\end{subfigure}
\hfill
\begin{subfigure}{.19\textwidth}
  \centering
  \includegraphics[width=\linewidth]{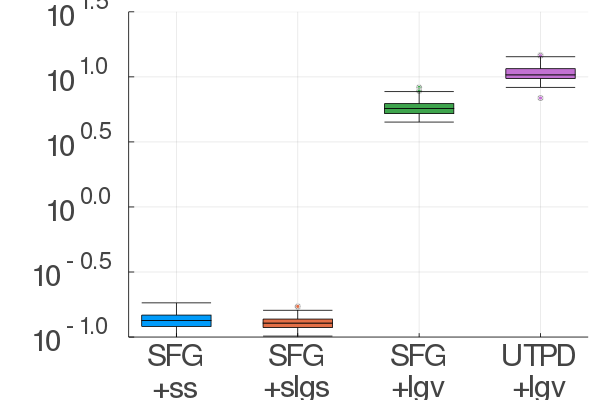}
  \caption{$d_x=12,p=15$}
\end{subfigure}
\hfill
\begin{subfigure}{.19\textwidth}
  \centering
  \includegraphics[width=\linewidth]{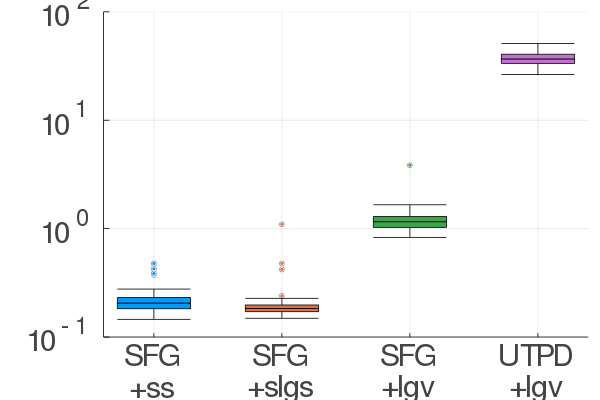}
  \caption{$d_x=15,p=16$}
\end{subfigure}
\caption{Distribution of Runtimes}
\label{fig:boxplots_runtime}
\end{figure*}

Table~\ref{table:average_optimal_compare} and Fig.~\ref{fig:boxplots_optval} report the optimal volumes obtained in our experiments, while Table~\ref{table:average_runtime_compare} and Fig.~\ref{fig:boxplots_runtime} report the observed runtimes.  Note that the vertical axis in all figures is logarithmic.

\subsection{Analysis}

Looking first at the three cases where $d_x = 3$, we see in Table~\ref{table:average_optimal_compare} that when $\countvecOp = d_x$ then the UTPD+lgv combination produces a noticeably better volume, but as we increase $\countvecOp$ the SFG+ss and SFG+lgv combinations match and then exceed the performance.  Interestingly SFG+slgs actually performs worse on the higher order cases.  We conjecture that this situation occurs because for higher orders there is more likely to be unproductive generators among those randomly chosen; allowing the scale factor for those generators to approach zero does not harm the ss or lgv objective functions but will significantly penalize the slgs objective function.

In contrast, as we increase $d_x$ we see a different trend.  UTPD+lgv consistently outperforms the others, particularly as the order of the zonotopes for SFG drops towards 1.  The SFG+slgs heuristic performs much closer to SFG+lgv than does the SFG+ss heuristic.

Examining the runtimes in Table~\ref{table:average_runtime_compare}, we see the expected behaviour for the SFG parameterizations: The ss and slgs heuristic objective functions are similarly fast, and the lgv objective function cost shows dramatic effects of both increasing $d_x$ and ${\countvecOp \choose d_x}$ (the latter explicitly computed for convenience in Table~\ref{table:combination_d_p_objective_compare}).  What is perhaps a little surprising is how the UTPD+lgv runtime is growing even faster than SFG+lgv as dimension increases, even though the objective function for the former is much cheaper to evaluate.  We attribute this effect to the fact that the UTPD parameterization has $\bigO(d_x^2)$ decision variables, compared to $\bigO(d_x)$ for the SFG parameterization.  Those extra decision variables produce much better zonotope volumes, but at a tremendous computational cost.

\subsection{Limitations}

The primary limitation of these experiments is that we considered a single algorithm to solve the single problem of finding a maximum volume zonotope invariant in a specified box constraint set over a fixed finite horizon for stable systems without inputs.  Going forward, we plan to compare these parameterizations and objective functions for more general types of invariant sets and multiple algorithms.

For this preliminary set of experiments we had quite limited compute resources and hence were unable to properly explore the effects of changing $d_x$ and $\countvecOp$ independently.  We also chose the extra generators in the SFG parameterization randomly, but there are better options available; for example~\cite{admioolam2017templatezonotope}. 

Finally, we considered only random dynamics in moderate dimensions.  It would be good to explore whether these techniques are at all feasible for real problems with dozens to hundreds of dimensions, such as those mentioned in~\cite{bak2017direct}.

\section{Conclusions and Future Works}

There has been considerable recent interest in using zonotopes for various forms of invariant and reachable sets in the past few years, and in several algorithms these zonotopes are constructed through optimizations.  In this paper we described two different parameterizations of zonotopes that have been used for such algorithms, and observed that the volume of the zonotope in each case is a log-concave function of the parameters and hence is amenable to efficient maximization.  For one particular invariant set algorithm, we explored four possible parameterization--objective function pairs: Both parameterizations with the true volume objective function, plus two heuristic objective functions for one of the parameterizations.  Perhaps not surprisingly the heuristics were fastest, but the quality of their results began to suffer in higher dimensions, while the relative quality and runtime of the more flexible parameterization both increased with dimension.

In future work we plan to expand the range of experiments in terms of dimension and order of the zonotopes, choice of template directions for the generators, the types of invariant sets being computed and the algorithms being used. 


\bibliographystyle{IEEEtran}
\bibliography{reference} 
\end{document}